\documentclass{procpavia}
\begin{document}
\pagestyle{prochead}

\title{Jastrow Two-nucleon Overlap Functions and Cross Sections of\\
$^{16}$O$(e,e^{\prime}NN)^{14}$C Reactions\footnote{The extended paper
was published in Phys. Rev. C \textbf{68}, 014617 (2003)}}

\author{D.~N.~Kadrev}
\affiliation{Institute for Nuclear Research and Nuclear Energy,
Bulgarian Academy of Sciences, Sofia 1784, Bulgaria}
\affiliation{Istituto Nazionale di Fisica
Nucleare, Sezione di Pavia, Pavia, Italy}

\author{M.~V.~Ivanov}
\affiliation{Institute for Nuclear Research and Nuclear Energy,
Bulgarian Academy of Sciences, Sofia 1784, Bulgaria}

\author{A.~N.~Antonov}
\affiliation{Institute for Nuclear Research and Nuclear Energy,
Bulgarian Academy of Sciences, Sofia 1784, Bulgaria}
\affiliation{Departamento de Fisica Atomica, Molecular y Nuclear,
Facultad de Ciencias Fisicas, Universidad Complutense de Madrid,
E-28040 Madrid, Espa\~na}

\author{C.~Giusti}
\affiliation{Istituto Nazionale di Fisica Nucleare, Sezione di Pavia,
Pavia, Italy}
\affiliation{Dipartimento di Fisica Nucleare e Teorica, Universit\`a di
Pavia, Pavia, Italy\\~\\}

\author{F.~D.~Pacati}
\affiliation{Istituto Nazionale di Fisica Nucleare, Sezione di Pavia,
Pavia, Italy}
\affiliation{Dipartimento di Fisica Nucleare e Teorica, Universit\`a di
Pavia, Pavia, Italy\\~\\}


\begin{abstract}
Using the relationship between the two-particle overlap functions
(TOF's) and the two-body density matrix (TDM), the TOF's for
$^{16}$O$(e,e^{\prime}pp)^{14}$C reaction are calculated on the basis
of TDM obtained with a Jastrow-type approach. The main contributions of
the removal of $^1S$ and $^3P$ $pp$-pairs from $^{16}$O are taken into
account in the calculations of the cross sections of the
$^{16}$O$(e,e^{\prime}pp)^{14}$C reaction using the Jastrow TOF's. The
contributions of the one-body and two-body delta currents are
considered. The results are compared with the calculations using TOF's
from other approaches.
\end{abstract}

\maketitle
\setcounter{page}{1}

\section{Introduction}

As known, two nucleons can be ejected from the nucleus by two-body
currents due to meson exchanges and delta-isobar excitation. But also,
the real or virtual photon can hit, through a one-body current either
nucleon of a correlated pair and both nucleons are then ejected
simultaneously from the nucleus. The role and relevance of these two
competing processes can be different in different reactions and
kinematics. It is thus possible to envisage situations where either
process is dominant and various specific effects can be disentangled
and separately investigated. This gives ground for studies of
short-range correlations (SRC) \cite{Gottfried,Oxford,Ant93,Mut+00} in
a nucleus by means of the two-nucleon knockout processes.

Various theoretical models for cross section calculations have been
developed in recent years in order to explore the effects of
ground-state $NN$ correlations on $(e,e^{\prime}NN)$
\cite{GP91,RVH+95,RVH+97,GP97,GPA+98,GPM+99} and $(\gamma,NN)$
\cite{GPR92,GP93,RVM+94,RMV+94,Ryc96,ic,GP98} knockout reactions. It
appears from these studies that the most promising tool for
investigating SRC in nuclei is represented by the $(e,e^{\prime}pp)$
reaction, where the effect of the two-body currents is less dominant as
compared to the $(e,e^{\prime}pn)$ and $(\gamma,NN)$ processes.
Measurements of the exclusive $^{16}$O$(e,e'pp)^{14}$C reaction
performed at NIKHEF in Amsterdam \cite{Ond+97,Ond+98,Ronald} and MAMI
in Mainz~\cite{Ros99,Ros00} have confirmed, in comparison with the
theoretical results, the validity of the direct knockout mechanism for
transitions to low-lying states of the residual nucleus and have given
clear evidence of SRC for the transition to the ground state of
$^{14}$C.

One of the main ingredients in the transition matrix elements of
exclusive two-nucleon knockout reactions is the two-nucleon overlap
function (TOF). The TOF contains information on nuclear structure and
correlations and allows one to write the cross section in terms of the
two-hole spectral function \cite{Oxford}. The TOF's and their
properties are widely reviewed, e.g., in \cite{BGP+85}.

In \cite{GP97} the TOF's for the $^{16}$O$(e,e'pp)^{14}$C reaction are
given by the product of a coupled and fully antisymmetrized pair
function of the shell model and a Jastrow-type correlation function
which incorporates SRC. A more sophisticated treatment is used in
\cite{GPA+98}, where the TOF's are obtained from an explicit
calculation of the two-proton spectral function of $^{16}$O
\cite{GAD+96}, which includes, with some approximations but
consistently, both SRC and long-range correlations (LRC).

A different method to calculate the TOF's has been suggested in
\cite{ADS+99} using the established general relationships connecting
TOF's with the ground state two-body density matrix (TDM). The
procedure is based on the asymptotic properties of the TOF's in
coordinate space, when the distance between two of the particles and
the center of mass of the remaining core becomes very large. This
procedure can be considered as an extension of the method suggested in
\cite{NWH93}, where the relationship between the one-body density
matrix and the one-nucleon overlap function is established. The latter
has been applied
\cite{SAD96,NDW97,NDD+96,DGA+97,GPD97,GPA00,IGA01,IGA02,GAI02} to
calculate the one-nucleon overlap functions, spectroscopic factors and
to make consistent calculations of the cross sections of different
one-nucleon removal reactions, such as $(p,d)$, $(e,e^{\prime}p)$, and
$(\gamma,p)$ \cite{SAD96,DGA+97,GPD97,GPA00,IGA01,IGA02,GAI02} on
$^{16}$O \cite{GPD97,GPA00,IGA02} and $^{40}$Ca \cite{IGA02,IGA01},
$(p,d)$ on $^{24}$Mg, $^{28}$Si and $^{32}$S \cite{GAI02}, as well as
$(e,e'p)$ on $^{32}$S \cite{GAI02} within various correlation methods.

The first aim of the present work is to apply the procedure suggested
in \cite{ADS+99} to calculate TOF's for $^{16}$O using the TDM
calculated in \cite{DKA+00} with the Jastrow correlation method (JCM),
which incorporates the nucleon-nucleon SRC. As a second aim, the
resulting two-proton overlap functions are used to calculate the cross
section of the $^{16}$O$(e,e^{\prime}pp)$ reaction for the transition
to the $0^+$ ground and the $1^+$ excited states of $^{14}$C. The cross
sections are calculated on the basis of the theoretical approach
developed in \cite{GP91,GP97,GPA+98}.

\section{Two-body density matrix and overlap functions} 

In this Section we present shortly the definitions and some properties
of the TDM and related quantities in the overlap function
representation. The method to extract the TOF's from the TDM
\cite{ADS+99} used in this work is also given.

The TDM is defined in coordinate space as:
\begin{equation}
\rho ^{(2)}(x_1,x_2;x_1^{\prime},x_2^{\prime})=\langle
{\Psi } ^{(A)}|a^{\dagger}(x_1) a^{\dagger}(x_2) a(x_2^{\prime})
a(x_1^{\prime})|{ \Psi }^{(A)}\rangle , \label{tdm}
\end{equation}
where $|{\Psi }^{(A)}\rangle $ is the antisymmetric $A$-fermion ground
state wave function normalized to unity and $a^{\dagger}(x)$, $a(x)$
are creation and annihilation operators at position $x$. The coordinate
$x$ includes the spatial coordinate $\mathbf{r}$ and spin and isospin
variables. The TDM $\rho ^{(2)}$ is trace-normalized to the number of
pairs of particles:
\begin{equation}
\mathrm{Tr} \; \rho ^{(2)}=\frac{1}{2} \int \rho ^{(2)}(x_1,x_2)
dx_1 dx_2 = \frac{A(A-1)}{2} . \label{norm2}
\end{equation}

Of direct physical interest is the decomposition of the TDM in
terms of the overlap functions between the $A$-particle ground
state and the eigenstates of the $(A-2)$-particle systems, since
TOF's can be probed in exclusive knockout reactions.

The TOF's are defined as the overlap between the ground state of the
target nucleus ${\Psi }^{(A)}$ and a specific state ${\Psi}
_{\alpha}^{(C)}$ of the residual nucleus ($C=A-2$) \cite{BGP+85}:
\begin{equation}
\Phi _{\alpha }(x_1,x_2)=\langle {\Psi }_{\alpha
}^{(C)}|a(x_1) a(x_2)|{\Psi }^{(A)}\rangle . \label{of2}
\end{equation}

Inserting a complete set of $(A-2)$ eigenstates $|\alpha (A-2)\rangle $
into Eq.~(\ref{tdm}) one gets
\begin{equation}
\rho ^{(2)}(x_1,x_2;x^{\prime}_1,x^{\prime}_2) =\sum
\limits_{\alpha } \Phi _{\alpha }^{*}(x_1,x_2) \Phi _{\alpha
}(x^{\prime}_1,x^{\prime}_2) . \label{ro_of}
\end{equation}

The norm of the two-body overlap functions defines the
spectroscopic factor
\begin{equation}
S_{\alpha }^{(2)}=\langle \Phi _{\alpha }|\Phi _{\alpha }\rangle .
\label{sf2}
\end{equation}

A procedure for obtaining the TOF's on the basis of the TDM
suggested in \cite{ADS+99} uses the particular asymptotic
properties of the TOF's.

In the case when two like nucleons (neutrons or protons) unbound to the
rest of the system are simultaneously transferred, the following
hyperspherical type of asymptotics is valid for the two-body overlap
functions \cite{BGP+85,Mer74,Ban80}
\begin{equation}
\Phi (r,R) \longrightarrow N \exp \left\{ -\sqrt{\frac{4m|E|}{\hbar
^{2}} \left( R^{2}+\frac{1}{4}r^{2}\right) }\right\} \left(
R^{2}+\frac{1}{4} r^{2}\right) ^{-5/2} , \label{of2as}
\end{equation}
where $r$ and $R$ are the magnitudes of the relative and center-of-mass
(c.m.) coordinates, $\mathbf{r}=\mathbf{r}_1-\mathbf{r}_2$ and
$\mathbf{R}=(\mathbf{r}_1+\mathbf{r} _2)/2$, respectively, $m$ is the
nucleon mass and $E=E^{(A)}-E^{(C)}$ is the two-nucleon separation
energy.

For a target nucleus with $J^{\pi }_{tar.}=0^{+}$ and for large
$r^{\prime}=a$ and $R^{\prime}=b$ a single term with $\nu_0$
(corresponding to the smallest two-nucleon separation energy) of
the radial part of the TOF $\Phi _{\nu_0 JSLlL_{R}}(r,R)$ can be
expressed in terms of the TDM as
\begin{equation}
\Phi _{\nu_0 JSLlL_{R}}(r,R) = \frac{\rho _{JSLlL_{R}}^{(2)}(r,R;a,b)
}{\Phi _{\nu_0 JSLlL_{R}}(a,b)} = \frac{\rho
_{JSLlL_{R}}^{(2)}(r,R;a,b)}{N \exp \left\{ -k\sqrt{\left(
b^{2}+\frac{1}{4}a^{2}\right) }\right\} \left( b^{2}+\frac{1}{4}
a^{2}\right) ^{-5/2}} , \label{p2of}
\end{equation}
where $k=(4m|E|/\hbar ^{2} )^{1/2}$ is constrained by the experimental
values of the two-nucleon separation energy $E$. The relationship
obtained in Eq.~(\ref{p2of}) makes it possible to extract TOF's with
quantum numbers $JSLlL_{R}$ from a given TDM. The coefficient $N$ and
the constant $k$ can be determined from the asymptotics of
${\rho^{(2)}_{JSLlL_{R}}}(r,R;r,R)$.

\section{Results}

\subsection{The two-proton overlap functions}

The procedure described briefly in Section 2 has been applied to
calculate the two-proton overlap functions in the $^{16}$O nucleus
for the transition to the $0^+$ ground and the $1^+$ excited
states of $^{14}$C. The TDM obtained in \cite{DKA+00} in the
framework of the low-order approximation (LOA) of the Jastrow
correlation method has been used \cite{Jas55}.

\begin{figure}
\begin{center}
\includegraphics[width=12cm]{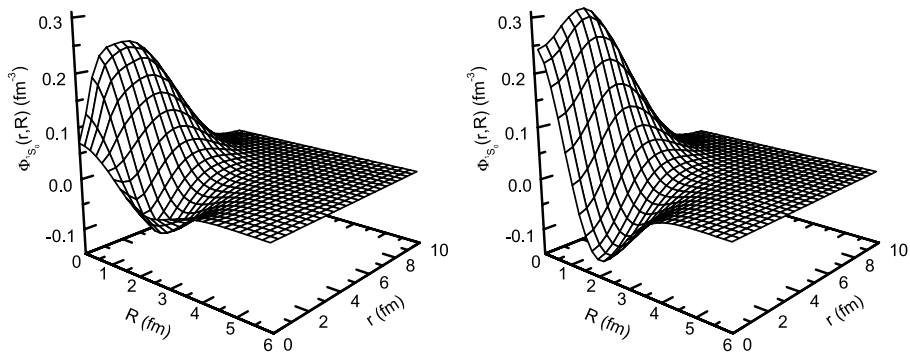}
\end{center} \caption{The $^1S_0$ two-proton overlap functions for
the nucleus $^{16}$O leading to the $0^+$ ground state of $^{14}$C
extracted from the JCM (left) and uncorrelated (right) two-body
density matrices. \label{f:1S0}}
\end{figure}

The TOF's for the $^1S_0$ and $^3P_1$ states are obtained in the
JCM. The result for $^1S_0$ state is presented in
Fig.~\ref{f:1S0}. It is compared with the uncorrelated TOF's
obtained applying the same procedure to the uncorrelated TDM. The
notation for the partial waves in our case is $^{2S+1}l_L$.

The spectroscopic factors corresponding to the $^1S_0$ and $^3P_1$
overlap functions are 0.958 and 0.957, respectively.

As a next step, we derive the total TOF ${\Phi}_{\nu JM}({x},{X})$ in
terms of a sum over all possible partial components, i.e.
\begin{equation}
{\Phi}_{\nu JM}({x},{X})=\sum_{LSlL_{R}}\Phi_{\nu JSLlL_{R}}(r,R)
A_{SLlL_{R}}^{JM }(\sigma _1,\sigma _2;\widehat{r}, \widehat{R}).
\label{total}
\end{equation}

We integrate the squared modulus of the total TOF in Eq.~(\ref{total})
over the angles and sum over the spin variables. The result can be
written in the form (for the smallest value of $\nu=\nu_0$):
\begin{equation}
\overline{|{\Phi}_{JM}({x},{X})|^{2}} {\equiv}
|\widetilde{\Phi}_{JM}({r},{R})|^{2} =\sum_{LSlL_R}
\rho^{(2)}_{JLSlL_R}(r,R), \label{21}
\end{equation}
where the bar denotes the integration over the angles and summation
over the spin variables, and $\widetilde{\Phi}_{JM}({r},{R})$ is the
radial part of the total TOF obtained after the integration and
summation. Using the asymptotics of $\widetilde{\Phi}_{JM}({r},{R})$ at
$r \longrightarrow a$, $R\longrightarrow b$ one can write:
\begin{equation}
\widetilde{\Phi}_{JM}({r},{R})= \frac{\displaystyle \sum_{LSlL_{R}}\rho
_{JSLlL_{R}}^{(2)}(r,R;a,b)}{N \exp \left\{ -k\sqrt{\left(
b^{2}+\frac{1}{4}a^{2}\right) }\right\} \left( b^{2}+\frac{1}{4}
a^{2}\right) ^{-5/2}}\ \ .\label{22}
\end{equation}

The results for the $^1S_0$ and $^3P_1$ partial components have a
similar behaviour as previous ones, the main difference is that
they are somewhat reduced in magnitude. The spectroscopic factor
corresponding to the total TOF is equal to unity in the
uncorrelated case and 0.965 in the Jastrow case.

The Jastrow TDM (including only SRC) is not ``rich'' enough to be able
to explain realistically transitions to all the excited states of
$^{14}$C. Therefore, only the transition to the 1$^+$ state is
considered in the present paper as an example of the applicability of
the method.

In the case of the transition to the $1^+$ excited state of
$^{14}$C $pp$-pairs in the states $^3P_{0,1,2}$ give main
contributions to the process. The value of the spectroscopic
factor e.g. for $^3P_{1}$ is 0.967 in the Jastrow case and unity
in the uncorrelated one.

\subsection{The $^{16}$O$(e,e^{\prime}pp)^{14}$C reaction}

The TOF's obtained from the TDM within the Jastrow correlation
method have been used to calculate the cross section of the
$^{16}$O$(e,e^{\prime}pp)^{14}$C knockout reaction in one-photon
exchange approximation \cite{Oxford,GP91}.

\begin{figure}
\begin{center}
\includegraphics[width=12cm]{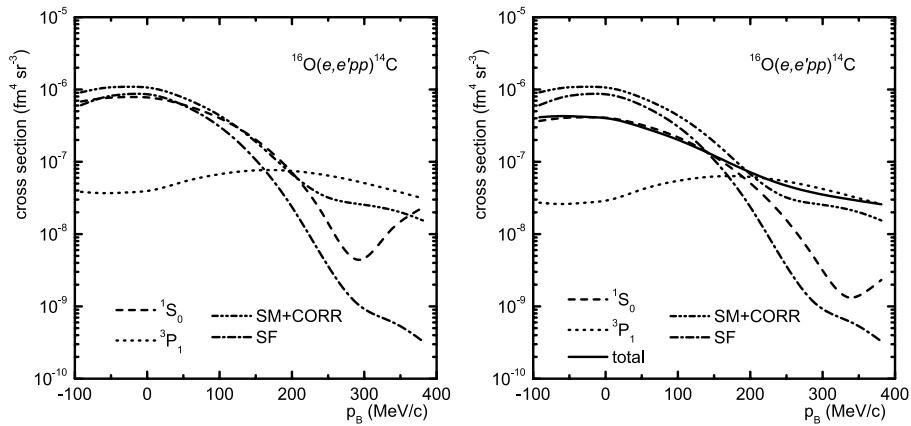}
\end{center} \caption{The differential cross section of the
$^{16}$O$(e,e^{\prime}p p)$ reaction as a function of the recoil
momentum $p_\mathrm{B}$ for the transition to the $0^+$ ground
state of $^{14}$C in the super-parallel kinematics with $E_0=855$
MeV, $\omega=215$ MeV and $q=316$ MeV/$c$. The curves are obtained
with different treatments of the TOF: $^1S_0$ (dashed line) and
$^3P_1$ (dotted line) as ``independent'' TOF's in the JCM in the
left panel and as partial components in the right panel.
\label{f:cs_sup}}
\end{figure}

As an example, the differential cross section calculated for the
transition to the $0^+$ ground state of $^{14}$C is shown in
Fig.~\ref{f:cs_sup} for the kinematical setting considered in the
experiment performed at MAMI \cite{Ros99,Ros00}.

The results are compared with the cross sections already shown in
\cite{GPA+98}, where the TOF is taken from a calculation of the
two-proton spectral function (SF) \cite{GAD+96}, where a two-step
procedure has been adopted to include both SRC and LRC.

In the figure are also shown for a comparison the results obtained
with a simpler approach, where the two-nucleon wave function is
given by the product of the pair function of the shell model and
of a Jastrow type central and state independent correlation
function (SM+CORR).

SRC are quite strong and even dominant for the $^1S_0$ state and
much weaker for the $^3P_1$ state. The role of the isobar current
is strongly reduced for $^1S_0$ $pp$ knockout, since there the
magnetic dipole $NN \leftrightarrow N\Delta$ transition is
suppressed \cite{GP98,delta}. As a consequence, in the figures the
$^1S_0$ results are dominated by the one-body current and thus by
SRC, while the $\Delta$ current gives the main contribution to
$^3P_1$ $pp$ knockout.

It can be seen from Fig.~\ref{f:cs_sup} that the cross section
calculated with the Jastrow TOF for the $^1S_0$ state is close to
the SF and also to the SM+CORR results at low values of
$p_\mathrm{B}$, up to $\sim 150-200$ MeV/$c$. For $p_\mathrm{B}
\geq 200$ MeV/$c$ $^3P_1$ knockout becomes dominant with all the
different treatments of the TOF. The results with the $^3P_1$ TOF
from the Jastrow TDM is however much larger than the SF result and
also larger than the SM+CORR cross section.

The cross section calculated with the total TOF, obtained from the
combination of the $^1S_0$ and $^3P_1$ partial components, are
shown in the right panel of Figs.~\ref{f:cs_sup}. In both
kinematical settings the $^1S_0$ component dominates at low values
of $p_\mathrm{B}$, while the $^3P_1$ component produces a strong
enhancement at high momenta.

Although obtained from a calculation of the TDM within the JCM
where only SRC are included, the TOF used in our calculations are
able to reproduce the main qualitative features which were found
in previous theoretical investigations. This means that the
procedure suggested in \cite{ADS+99} to calculate the TOF's from
the TDM can be applied and exploited in the study of two-nucleon
knockout reactions.

\begin{figure}
\begin{center}
\includegraphics[width=6cm]{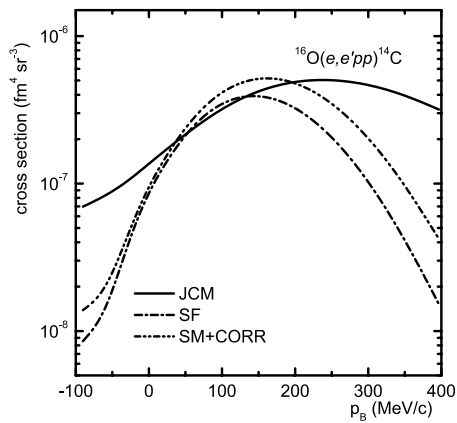}
\end{center}
\caption{The differential cross section of the
$^{16}$O$(e,e^{\prime}p p)$ reaction as a function of the recoil
momentum $p_\mathrm{B}$ for the transition to the $1^+$ excited
state of $^{14}$C in the same kinematics as in
Fig.~\ref{f:cs_sup}. \label{f:cs_sup1+}}
\end{figure}

The differential cross section calculated for the transition to
the $1^+$ state of $^{14}$C, at 11.3 MeV excitation energy, is
shown in Fig.~\ref{f:cs_sup1+} in the same kinematical setting
already considered for the $0^+$ state in Fig.~\ref{f:cs_sup}.
With respect to the other results, the Jastrow TOF produces in the
super-parallel kinematics a strong enhancement at high momenta,
which makes the shape of the cross sections larger and flatter
than that with the SF and SM+CORR TOF's.

\section{Conclusions}

The results of the present work can be summarized as follows:

\begin{itemize} \item[i)] The two-nucleon overlap functions (and their
norms, the spectroscopic factors) corresponding to the knockout of
two protons from the ground state of $^{16}$O and the transition
to the ground and $1^+$ excited states of $^{14}$C are calculated
using the recently established relationship \cite{ADS+99} between
the TOF's and the TDM. In the calculations the TDM obtained within
the JCM \cite{DKA+00} is used. Though only SRC are accounted for
in the Jastrow TDM, the results can be considered as a first
attempt to use an approach which fulfils the general necessity the
TOF's to be extracted from theoretically calculated TDM's
corresponding to realistic wave functions of the nuclear states.

\item[ii)] The TOF's extracted from the Jastrow TDM are included in
the theoretical approach of \cite{GP91,GP97,GPA+98} to calculate
the cross section of the $^{16}$O$(e,e^{\prime}pp)^{14}$C knockout
reaction. Numerical results in different kinematics are compared
with the cross sections calculated, within the same theoretical
model for the reaction mechanism, with different treatments of the
TOF, in particular with the more refined approach of
\cite{GPA+98,GAD+96}, where the TOF's are obtained from a
calculation of the two-proton spectral function of $^{16}$O where
both SRC and LRC are included. The cross sections calculated in
the present work, where the TOF's are extracted from the Jastrow
TDM, confirm the dominant contribution of $^{1}S_0$ $pp$ knockout
at low values of recoil momentum, up to $\simeq 150-200$ MeV/$c$.
The $^{3}P_1$ contribution is mainly responsible for the
high-momentum part of the cross section at $p_\mathrm{B} \geq 200$
MeV/$c$.

\item[iii)] Our method is applied in the present work only to the $0^+$
ground and the $1^+$ excited states of $^{14}$C. The main aim was to
check the practical application of all steps of the method to a given
state of the residual nucleus. Therefore, the results obtained for the
$^{16}$O$(e,e^{\prime}pp)^{14}$C reaction, which are able to reproduce
the main qualitative features of the cross sections calculated with
different treatments of the TOF's, can serve as an indication of the
reliability of the method, that can be applied to a wider range of
situations and, as an alternative to an explicit calculation of the
two-hole spectral function, to more refined approaches of the TDM
\cite{PWP92,Hei+99,Co+00,Mut+00}. \end{itemize}

\section*{Acknowledgments}

One of the authors (D.~N.~K.) would like to thank the Pavia Section of
the INFN for the warm hospitality and for providing the necessary
fellowship. A.N.A. is grateful for support during his stay at the
Complutense University of Madrid to the State Secretariat of Education
and Universities of Spain (N/Ref. SAB2001-0030). Three of the authors
(A.N.A., M.V.I. and D.N.K.) are thankful to the Bulgarian National
Science Foundation for partial support under the Contracts Nos.
$\Phi$-809 and $\Phi$-905.

\end{document}